\documentclass[english,twocolumn]{IEEEtran}
\usepackage[T1]{fontenc}
\usepackage[latin9]{inputenc}
\usepackage{color}
\usepackage{mathrsfs}
\usepackage{amsthm}
\usepackage{amsmath}
\usepackage{amssymb}
\usepackage{esint}

\makeatletter
\theoremstyle{definition}
\newtheorem{assumption}{Assumption}
  \theoremstyle{remark}
  \newtheorem{rem}{\protect\remarkname}
  \theoremstyle{plain}
  \newtheorem{thm}{\protect\theoremname}

\usepackage{cite}
\allowdisplaybreaks
\IEEEoverridecommandlockouts

\makeatother

\usepackage{babel}
\providecommand{\remarkname}{Remark}
\providecommand{\theoremname}{Theorem}

\begin{document}

\title{Time-Varying Input and State Delay Compensation for Uncertain Nonlinear
Systems%
\thanks{Rushikesh Kamalapurkar, Nicholas Fischer, Serhat Obuz, and Warren
E. Dixon are with the Department of Mechanical and Aerospace Engineering,
University of Florida, Gainesville, FL, USA. Email: rkamalapurkar@ufl.edu,
nic.r.fischer@gmail.com, \{serhat.obuz, wdixon\}@ufl.edu.%
}%
\thanks{This research is supported in part by NSF award numbers 1161260 and
1217908, ONR grant number N00014-13-1-0151, and a contract with the
AFRL Mathematical Modeling and Optimization Institute. Any opinions,
findings and conclusions or recommendations expressed in this material
are those of the authors and do not necessarily reflect the views
of the sponsoring agency.%
}}

\author{Rushikesh Kamalapurkar, Nicholas Fischer, Serhat Obuz, and Warren
E. Dixon}
\maketitle
\begin{abstract}
A robust controller is developed for uncertain, second-order nonlinear
systems subject to simultaneous unknown, time-varying state delays
and known, time-varying input delays in addition to additive, sufficiently
smooth disturbances. An integral term composed of previous control
values facilitates a delay-free open-loop error system and the development
of the feedback control structure. A stability analysis based on Lyapunov-Krasovskii
(LK) functionals guarantees uniformly ultimately bounded tracking
under the assumption that the delays are bounded and slowly varying.
\end{abstract}

\section{Introduction\label{sec:Introduction}}

Numerous control techniques exist for linear systems with constant
input delays (cf. \cite{Krstic2009b,Chiasson2007,Gu2003} and references
therein). Many of these results are extensions of classic Smith predictors
\cite{Smith1959}, Artstein model reduction \cite{Artstein1982},
or finite spectrum assignment \cite{Manitius1979}. Results that focus
on simultaneous constant state and input delays for linear systems
are provided in \cite{Bekiaris-Liberis.Krstic2010,Jankovic2009,Jankovic2010}.
Extensions of linear control techniques to time-varying input delays
are also available \cite{Nihtila1989,Lozano2004,Yue2005a,Wang2007,Richard2003,Krstic2010a}.

For nonlinear systems, controllers considering constant \cite{Ge2004,Ge2005,Yoo2007,Hua2007,Wu2009,Wang2009a,Tong2010,Kuperman2011}
and time-varying \cite{Huang2007,Wu2009,Ren2009,Yoo2009,Wang2010a,Niu2005,Chen2010,Mirkin2010,Mirkin2011,Sharma2012,Fischer.Kamalapurkar.ea2012b}
state delays have been recently developed. However, linear results
considering delayed inputs are far less prevalent, especially for
systems with model uncertainties and/or disturbances. Examples of
these include constant input delay results in \cite{Mazenc.Niculescu2011,Karafyllis2011,Krstic2008b,Bresch-Pietri2009,Mazenc2004,Carravetta.Palumbo.ea2010,Chen2008b,Mazenc.Niculescu.ea2012,Pepe.Jiang.ea2008,Krstic2010,Sharma2011,Castillo-Toledo2010,Obuz.Tatlicioglu.ea2012}
and time-varying input delay results based on LMI conditions \cite{Jiao.Yang.ea2011,Liu.Zhang.ea2011},
backstepping \cite{Karafyllis2006,Bekiaris-Liberis2011,Bekiaris-Liberis.Krstic2012}
and other robust techniques \cite{Fischer.Kamalapurkar.ea2012a}.
Even more unique are results that consider both state and input delays
in nonlinear systems. Recently in \cite{Bekiaris-Liberis.Krstic2012},
the predictor-based techniques in \cite{Bekiaris-Liberis.Krstic2010}
were extended to nonlinear systems with time-varying delays in the
state and/or the input utilizing a backstepping transformation to
construct a predictor-based compensator. The development in \cite{Bekiaris-Liberis.Krstic2012}
requires knowledge of the plant dynamics and assumes that the plant
is disturbance-free.

In this paper, we expand our previous time-varying input delay result
\cite{Fischer.Kamalapurkar.ea2012a} in two directions: a) Utilizing
techniques for constant input-delayed systems first introduced in
\cite{Obuz.Tatlicioglu.ea2012}, we consider time-varying input delays
in a nonlinear plant, and b) we add the ability to compensate for
simultaneous arbitrarily large unknown time-varying state delays based
on the techniques in \cite{Fischer.Kamalapurkar.ea2012b}. Robust
control methods are developed to compensate for the unknown time-varying
state delays. To compensate for the input delay, an integral term
composed of previous control values is used to yield a delay-free
open-loop system. A Lyapunov-based stability analysis motivated by
Lyapunov-Krasovskii (LK) functionals demonstrates the ability to achieve
uniformly ultimately bounded tracking in the presence of model uncertainty,
additive sufficiently smooth disturbances, and simultaneous time-varying
state and input delays. The result is based on the assumption that
the unknown state delay is bounded and slowly varying. Improving on
the result in \cite{Fischer.Kamalapurkar.ea2012a}, we relax previous
sufficient conditions on the control gains that required knowledge
of the second derivative of the input delay.

\section{Dynamic System\label{sec:Dynamic-System}}

Consider a class of second-order (Euler-Lagrange-like) nonlinear systems
given by:%
\footnote{The result in this paper can be extended to $n^{th}$-order nonlinear
systems following a similar development to that presented in \cite{Sharma2012}.%
}%
\footnote{For notational brevity, unless otherwise specified, the domain of
all the functions is assumed to be $\mathbb{R}_{\geq0}$. Furthermore,
unless otherwise specified, all mathematical quantities are assumed
to be time-varying and time-dependence is suppressed in equations
and definitions. For example, the trajectory $x:\mathbb{R}_{\geq0}\to\mathbb{R}^{n}$
is defined by abuse of notation as $x\in\mathbb{R}^{n}$ and unless
otherwise specified, an equation (inequality) of the form $f+h\left(y,t\right)=\left(\leq\right)g\left(x\right)$
is interpreted as $f\left(t\right)+h\left(y\left(t\right),t\right)=\left(\leq\right)g\left(x\left(t\right)\right)$
for all $t\in\mathbb{R}_{\geq0}$.%
}
\begin{equation}
\ddot{x}=f\left(x,\dot{x},t\right)+g\left(x\left(t-\tau_{s}\right),\dot{x}\left(t-\tau_{s}\right),t\right)+d+u_{\tau_{i}},\label{eq: dynamics}
\end{equation}
where $x,\dot{x}\in\mathbb{R}^{n}$ are the system states, $u\in\mathbb{R}^{n}$
is the control input, $f:\mathbb{R}^{n}\times\mathbb{R}^{n}\times\left[0,\infty\right)\rightarrow\mathbb{R}^{n}$
is an unknown function, uniformly bounded in $t$, $g:\mathbb{R}^{n}\times\mathbb{R}^{n}\times\left[0,\infty\right)\rightarrow\mathbb{R}^{n}$
is an unknown function with delayed internal state, uniformly bounded
in $t$, $d\in\mathbb{R}^{n}$ denotes a sufficiently smooth disturbance
(e.g., unmodeled effects), and $\tau_{i},\tau_{s}\in\left[0,\infty\right)$
denote time-varying, non-negative input and state delays, respectively.

The subsequent development is based on the assumption that $x$ and
$\dot{x}$ are measurable outputs. Throughout the paper, a time-dependent
delayed function is denoted as
\[
\zeta_{\tau}\triangleq\begin{cases}
\zeta\left(t-\tau\right) & t-\tau>t_{0}\\
0 & t-\tau\leq t_{0},
\end{cases}
\]
 where $t_{0}$ denotes the initial time. Thus, $u_{\tau_{i}}$ is
defined by 
\[
u_{\tau_{i}}\triangleq\begin{cases}
u\left(t-\tau_{i}\right) & t-\tau_{i}>t_{0}\\
0 & t-\tau_{i}\leq t_{0}.
\end{cases}
\]
Additionally, $\left\Vert \cdot\right\Vert $ denotes the Euclidean
norm of a vector and the following assumptions will be exploited.
\begin{assumption}
\label{ass: fg}Each of the functions $f$ and $g$, along with their
first and second partial derivatives, is bounded on each subset of
its domain of the form $K\times\left[0,\infty\right)$, where $K\subset\mathbb{R}^{n}\times\mathbb{R}^{n}$
is compact. Furthermore, given such compact $K,$ the corresponding
bound is known. 
\end{assumption}

\begin{assumption}
\label{ass: disturbance}The nonlinear disturbance term and its time
derivative are bounded by known constants.%
\footnote{Many practical disturbance terms are continuous including friction
(see \cite{Makkar2007,Makkar2005}), wind disturbances, wave/ocean
disturbances, etc.%
}
\end{assumption}

\begin{assumption}
\label{ass: desired-traj}The desired trajectory $x_{d}\in\mathbb{R}^{n}$
is designed such that $x_{d}^{\left(i\right)}\in\mathbb{R}^{n},\:\forall i=0,1,...,3$
exist and are bounded by known positive constants, where the superscript
$\left(i\right)$ denotes the $i^{th}$ time derivative.%
\footnote{\noindent Many guidance and navigation applications utilize smooth,
high-order differentiable desired trajectories. Curve fitting methods
can also be used to generate sufficiently smooth time-varying trajectories.%
}
\end{assumption}

\begin{assumption}
\label{ass: delay bounds}The input and state delays are bounded such
that $0\leq\tau_{i}\leq\varphi_{i_{1}}$ and $0\leq\tau_{s}\leq\varphi_{s_{1}},$
and the rate of change of the delays are bounded such that $\left|\dot{\tau}_{i}\right|\leq\varphi_{i_{2}}<1$
and $\left|\dot{\tau}_{s}\right|\leq\varphi_{s_{2}}<1$ where $\varphi_{j}\in\mathbb{R}^{+}\:\forall j=i_{1},i_{2},s_{1},s_{2}$
are known constants. Furthermore, the bounds on the input delay satisfy
$\varphi_{i_{1}}+\varphi_{i_{2}}<1$. The state delay is assumed to
be unknown, while the input delay is assumed to be known.\end{assumption}
\begin{rem}
In Assumption \ref{ass: delay bounds}, the slowly time-varying constraint
(i.e., $\left|\dot{\tau}_{i,s}\right|\leq\varphi_{i_{2},s_{2}}<1$)
is common to results which utilize classical LK functionals to compensate
for time-varying time-delays \cite{Richard2003}. Knowledge of the
state delays in the system is not required; however, the input delays
present a more significant challenge. Although the controller requires
the input delay to be known so that the interval of past control values
can be properly sized, numerical simulations illustrate robustness
to uncertainties in the input delay.
\end{rem}

\section{Control Objective\label{sec:Control-Objective}}

The objective is to design a controller that will ensure the system
state $x$ of the system in (\ref{eq: dynamics}) tracks a desired
state trajectory. To quantify the control objective, a tracking error,
denoted by $e_{1}\in\mathbb{R}^{n}$, is defined as
\begin{equation}
e_{1}\triangleq x_{d}-x.\label{eq: e1}
\end{equation}
To facilitate the subsequent analysis, two auxiliary tracking errors
$e_{2},r\in\mathbb{R}^{n}$ are defined as \cite{Obuz.Tatlicioglu.ea2012}
\begin{align}
e_{2} & \triangleq\dot{e}_{1}+\alpha_{1}e_{1},\label{eq: e_2}\\
r & \triangleq\dot{e}_{2}+\alpha_{2}e_{2}+e_{u},\label{eq: r}
\end{align}
where $\alpha_{1},\alpha_{2}\in\mathbb{R}$ denote constant positive
control gains, and $e_{u}\in\mathbb{R}^{n}$ denotes the mismatch
between the delayed control input and the computed control input,
defined as%
\footnote{Let $h\triangleq\max\left(t_{0},t-\tau_{i}\right)$. Then, $h:[0,\infty)\to[0,\infty)$
is continuous. Further, since $u\left(t_{0}\right)=0$, $u_{\tau_{i}}=u\left(h\right)$,
and $e_{u}=u\left(h\right)-u$. Hence, $e_{u}$ is a continuous function
of time if $u$ is a continuous function of time, and $e_{u}\left(t_{0}\right)=0$.%
}
\begin{equation}
e_{u}\triangleq u_{\tau_{i}}-u.\label{eq: e_u}
\end{equation}
The auxiliary signal $e_{u}$ injects a delay-free control input into
the error system development. In contrast to the development in \cite{Fischer.Kamalapurkar.ea2012a},
the term in (\ref{eq: e_u}) is embedded in a higher order derivative
(i.e., $r$ instead of $e_{2}$). Functionally, $e_{u}$ still injects
an integral of past control values into the open-loop system; however,
the development introduces fewer cross-terms. The auxiliary signal
$r$ is introduced to facilitate the stability analysis and is not
used in the control design since the expression in (\ref{eq: r})
depends on the unmeasurable state $\ddot{x}$. The structure of the
error systems is motivated by the need to inject and cancel terms
in the subsequent stability analysis as demonstrated in Section \ref{sec:Stability-Analysis}. 

Using (\ref{eq: dynamics}), (\ref{eq: e1}), (\ref{eq: e_2}), and
(\ref{eq: e_u}) to eliminate the delayed input term, (\ref{eq: r})
can be represented as 
\begin{equation}
r=S_{1}+S_{2}-u,\label{eq: open loop}
\end{equation}
where the auxiliary functions $S_{1}\in\mathbb{R}^{n}$ and $S_{2}\in\mathbb{R}^{n}$
are defined as  
\begin{align*}
S_{1} & \triangleq f\left(x_{d},\dot{x}_{d},t\right)-f\left(x,\dot{x},t\right)+g\left(x_{d\tau_{s}},\dot{x}_{d\tau_{s}},t\right)\\
 & -g\left(x_{\tau_{s}},\dot{x}_{\tau_{s}},t\right)+\alpha_{1}\dot{e}_{1}+\alpha_{2}e_{2},\\
S_{2} & \triangleq\ddot{x}_{d}-f\left(x_{d},\dot{x}_{d},t\right)-g\left(x_{d\tau_{s}},\dot{x}_{d\tau_{s}},t\right)-d.
\end{align*}

Based on (\ref{eq: open loop}) and the subsequent stability analysis,
\begin{equation}
u\triangleq\left(k_{s}+1\right)\left(e_{2}-e_{2}\left(t_{0}\right)\right)+v,\label{eq: u}
\end{equation}
where $v\in\mathbb{R}^{n}$ is the solution to the following differential
equation
\begin{equation}
\dot{v}=\left(k_{s}+1\right)\left(\alpha_{2}e_{2}+e_{u}\right),\: v\left(t_{0}\right)=0,\label{eq: v}
\end{equation}
and $k_{s}\in\mathbb{R}$ is a positive constant control gain.

The closed-loop error system can be developed by taking the time derivative
of (\ref{eq: open loop}) and substituting for (\ref{eq: r}) and
the time derivative of (\ref{eq: u}) to yield
\begin{equation}
\dot{r}=\tilde{N}+N_{d}-e_{2}-\left(k_{s}+1\right)r,\label{eq: closed loop}
\end{equation}
where $\tilde{N}\in\mathbb{R}^{n}$ and $N_{d}\in\mathbb{R}^{n}$
are defined as
\begin{align}
\tilde{N} & \triangleq\dot{S}_{1}+e_{2},\label{eq: N tilde}\\
N_{d} & \triangleq\dot{S}_{2}.\label{eq: Nd}
\end{align}
The control design in (\ref{eq: u}) and (\ref{eq: v}) is motivated
by the desire to eliminate the delayed input, yielding the closed-loop
error system in (\ref{eq: closed loop}). The structure of (\ref{eq: closed loop})
is advantageous because it facilitates the stability analysis by segregating
terms that can be upper bounded by a state-dependent term and terms
that can be upper bounded by constants. Based on Assumptions \ref{ass: disturbance}
and \ref{ass: desired-traj}, the following inequalities can be developed
from the expression in (\ref{eq: Nd}):
\begin{equation}
\left\Vert N_{d}\right\Vert \leq\zeta_{N_{d1}},\label{eq: Nd_bound}
\end{equation}
where $\zeta_{N_{d1}}\in\mathbb{R},$ is a known positive constant.
The Mean Value Theorem can be utilized to find an upper bound for
the expression in (\ref{eq: N tilde}) as \cite[Appendix A]{arxivKamalapurkar.Klotz.ea2013}
\begin{equation}
\left\Vert \tilde{N}\right\Vert \leq\rho_{1}\left(\left\Vert z\right\Vert \right)\left\Vert z\right\Vert +\rho_{2}\left(\left\Vert z_{\tau_{s}}\right\Vert \right)\left\Vert z_{\tau_{s}}\right\Vert ,\label{eq: N tilde bound}
\end{equation}
where $z\in\mathbb{R}^{4n}$ denotes the vector
\begin{equation}
z\triangleq\left[e_{1}^{T},\:\: e_{2}^{T},\:\: r^{T},\:\: e_{u}^{T}\right]^{T},\label{eq: z}
\end{equation}
and the bounding terms $\rho_{1},\rho_{2}:[0,\infty)\to[0,\infty)$
are positive, non-decreasing and radially unbounded functions.%
\footnote{For some classes of systems, the bounding functions $\rho_{1}$ and
$\rho_{2}$ could be selected as constants. For these classes of systems,
a global uniformly ultimately bounded result can be obtained as described
in Remark \ref{rem:rho}.%
} The upper bound for the auxiliary function $\tilde{N}$ is segregated
into delay-free and delay-dependent bounding functions to eliminate
the delayed terms with the use of an LK term in the stability analysis. 

To facilitate the subsequent stability analysis, several auxiliary
terms are introduced. Let $\rho:[0,\infty)\rightarrow[0,\infty)$
be an auxiliary bounding function defined as 
\begin{equation}
\rho\left(\left\Vert z\right\Vert \right)=\sqrt{\left(\gamma_{1}+2\gamma_{2}\varphi_{s_{1}}\right)\rho_{2}^{2}\left(\left\Vert z\right\Vert \right)+3\rho_{1}^{2}\left(\left\Vert z\right\Vert \right)},\label{eq:rho}
\end{equation}
where $\gamma_{1}$ and $\gamma_{2}$ are positive adjustable constants,
and let $\acute{z}\in\mathbb{R}^{3n}$ be defined as
\begin{equation}
\acute{z}\triangleq\left[e_{1}^{T},\:\: e_{2}^{T},\:\: r^{T}\right]^{T}.\label{eq: z-1}
\end{equation}
Auxiliary bounding constants $\sigma,\delta\in\mathbb{R}$ are defined
as
\begin{align}
\sigma & \triangleq\frac{1}{2}\min\left\{ \frac{\alpha_{1}}{2},\frac{\alpha_{2}}{2},1,\frac{\omega\left(1-\varphi_{i_{2}}\right)}{6\varphi_{i_{1}}}\right\} ,\label{eq: sigma}\\
\delta & \triangleq\frac{1}{2}\min\Biggl\{\sigma,\frac{\omega\left(1-\varphi_{i_{2}}\right)}{3\varphi_{i_{1}}},\frac{\left(1-\varphi_{i_{2}}\right)}{3\varphi_{i_{1}}},\nonumber \\
 & \frac{\gamma_{2}\left(1-\varphi_{s_{2}}\right)}{\gamma_{1}},\frac{\left(1-\varphi_{s_{2}}\right)}{2\varphi_{s_{1}}}\Biggl\},\label{eq:delta}
\end{align}
where $\omega\in\mathbb{R}$ is a known, positive, adjustable constant. 

Let 
\[
\mathscr{D}\triangleq\left\{ \xi\in\mathbb{R}^{3n+4}|\left\Vert \xi\right\Vert <\inf\left\{ \rho^{-1}\left(\left[\sqrt{2k_{s}\sigma},\infty\right)\right)\right\} \right\} ,
\]
and 
\[
\mathcal{S}_{\mathscr{D}}\triangleq\left\{ \xi\in\mathscr{D}|\left\Vert \xi\right\Vert <\sqrt{\frac{1}{2}}\inf\left\{ \rho^{-1}\left(\left[\sqrt{2k_{s}\sigma},\infty\right)\right)\right\} \right\} ,
\]
where, for a set $A\subset\mathbb{R}$, the inverse image $\rho^{-1}\left(A\right)\subset\mathbb{R}$
is defined as $\rho^{-1}\left(A\right)\triangleq\left\{ a\in\mathbb{R}\mid\rho\left(a\right)\in A\right\} $.
Furthermore, let the functions%
\footnote{The construction of $P_{LK}$, $Q_{LK}$, $R_{LK}$, and $S_{LK}$
is based on LK functionals. However, in this result, they are to be
interpreted as time-varying signals that are a part of the system
state.%
} $P_{LK}:[0,\infty)\to[0,\infty),$ $Q_{LK}:[0,\infty)\to[0,\infty),$
$R_{LK}:[0,\infty)\to[0,\infty)$, and $S_{LK}:[0,\infty)\to[0,\infty)$
be defined as 
\begin{align}
P_{LK} & \triangleq\varphi_{i_{1}}\int_{t-\tau_{i}}^{t}\left\Vert \dot{u}\left(\theta\right)\right\Vert ^{2}d\theta,\label{eq: PLK}\\
Q_{LK} & \triangleq\omega\int_{t-\tau_{i}}^{t}\left(\int_{s}^{t}\left\Vert \dot{u}\left(\theta\right)\right\Vert ^{2}d\theta\right)ds,\label{eq: QLK}\\
R_{LK} & \triangleq\frac{\gamma_{1}}{2k_{s}}\int_{t-\tau_{s}}^{t}\rho_{2}^{2}\left(\left\Vert z\left(\sigma\right)\right\Vert \right)\left\Vert z\left(\sigma\right)\right\Vert ^{2}d\sigma,\label{eq: RLK}\\
S_{LK} & \triangleq\frac{\gamma_{2}}{k_{s}}\int_{t-\tau_{s}}^{t}\left(\int_{s}^{t}\rho_{2}^{2}\left(\left\Vert z\left(\sigma\right)\right\Vert \right)\left\Vert z\left(\sigma\right)\right\Vert ^{2}d\sigma\right)ds.\label{eq: SLK}
\end{align}
Additionally, let $y\in\mathbb{R}^{3n+4}$ be defined as
\begin{equation}
y\triangleq\left[\begin{array}{ccccc}
\acute{z}^{T} & \sqrt{P_{LK}} & \sqrt{Q_{LK}} & \sqrt{R_{LK}} & \sqrt{S_{LK}}\end{array}\right]^{T}.\label{eq: y}
\end{equation}

\section{Stability Analysis\label{sec:Stability-Analysis}}
\begin{thm}
\noindent \label{thm: thm}Given the dynamics in (\ref{eq: dynamics}),
provided the control gains are selected based on the following sufficient
conditions
\begin{gather}
\alpha_{1}>1,\:\alpha_{2}>2,\:\gamma_{1}>\frac{1}{\left(1-\varphi_{s_{2}}\right)},\:\omega>\frac{3\varphi_{i_{1}}}{\left(1-\varphi_{i_{2}}\right)},\label{eq: gain cond}
\end{gather}
and the input delay is small enough so that there exists a gain $k_{s}$
that satisfies%
\footnote{Since $\delta$ increases with increasing $k_{s}$, the left-hand
side of (\ref{eq:DelayCond2}) decreases with increasing $k_{s}$.
Since $\rho$ is a nondecreasing function, the right-hand side of
(\ref{eq:DelayCond2}) is nondecreasing with respect to $k_{s}$.
Hence, (\ref{eq:DelayCond2}) can be satisfied for some $k_{s}$.
Furthermore, for any given $k_{s}$, (\ref{eq:DelayCond1}) is satisfied
if the delay is small enough.%
} 
\begin{align}
\varphi_{i_{1}} & <\frac{k_{s}}{6\left(\omega+1\right)\left(k_{s}+1\right)^{2}},\label{eq:DelayCond1}\\
\frac{3\zeta_{N_{d1}}^{2}}{k_{s}\delta} & <\left(\inf\left\{ \rho^{-1}\left(\left[\sqrt{2k_{s}\sigma},\infty\right)\right)\right\} \right)^{2},\label{eq:DelayCond2}
\end{align}
the controller given in (\ref{eq: u}) and (\ref{eq: v}) ensures
uniformly ultimately bounded tracking in the sense that $\lim\sup_{t\to\infty}\left\Vert y\right\Vert \leq\sqrt{\frac{3\zeta_{N_{d1}}^{2}}{k_{s}\delta}}$,
provided $y\left(t_{0}\right)\in\mathcal{S}_{\mathscr{D}}$.\end{thm}
\begin{IEEEproof}
Let $V:\mathscr{D}\rightarrow\mathbb{R}$ be a candidate Lyapunov
function defined as
\begin{multline}
V\triangleq\frac{1}{2}e_{1}^{T}e_{1}+\frac{1}{2}e_{2}^{T}e_{2}+\frac{1}{2}r^{T}r+P_{LK}+Q_{LK}\\
+R_{LK}+S_{LK},\label{eq: V}
\end{multline}
which satisfies the following inequalities:
\begin{equation}
\frac{1}{2}\left\Vert y\right\Vert ^{2}\leq V\left(y\right)\leq\left\Vert y\right\Vert ^{2}.\label{eq: V bound}
\end{equation}
The time derivative of (\ref{eq: V}) can be found by applying the
Leibniz Rule to (\ref{eq: PLK}), (\ref{eq: QLK}), (\ref{eq: RLK})
and (\ref{eq: SLK}), and by substituting (\ref{eq: e1})-(\ref{eq: r}),
(\ref{eq: u}), and (\ref{eq: closed loop}), yielding
\begin{align}
\dot{V} & =e_{1}^{T}\left(e_{2}-\alpha_{1}e_{1}\right)+e_{2}^{T}\left(r-\alpha_{2}e_{2}-e_{u}\right)\nonumber \\
 & +r^{T}\left(\tilde{N}+N_{d}-e_{2}-\left(k_{s}+1\right)r\right)\nonumber \\
 & +\left(\omega\tau_{i}+\varphi_{i_{1}}\right)\left(k_{s}+1\right)^{2}\left\Vert r\right\Vert ^{2}-\varphi_{i_{1}}\left(1-\dot{\tau}_{{\color{blue}i}}\right)\left\Vert \dot{u}_{\tau_{i}}\right\Vert ^{2}\nonumber \\
 & -\omega\left(1-\dot{\tau}_{i}\right)\int_{t-\tau_{i}}^{t}\left\Vert \dot{u}\left(\theta\right)\right\Vert ^{2}d\theta\nonumber \\
 & +\left(\frac{\gamma_{1}}{2k_{s}}+\frac{\gamma_{2}}{k_{s}}\tau_{s}\right)\rho_{2}^{2}\left(\left\Vert z\right\Vert \right)\left\Vert z\right\Vert ^{2}\nonumber \\
 & -\frac{\gamma_{1}\left(1-\dot{\tau}_{s}\right)}{2k_{s}}\rho_{2}^{2}\left(\left\Vert z_{\tau_{s}}\right\Vert \right)\left\Vert z_{\tau_{s}}\right\Vert ^{2}\nonumber \\
 & -\frac{\gamma_{2}}{k_{s}}\left(1-\dot{\tau}_{s}\right)\int_{t-\tau_{s}}^{t}\rho_{2}^{2}\left(\left\Vert z\left(\theta\right)\right\Vert \right)\left\Vert z\left(\theta\right)\right\Vert ^{2}d\theta.\label{eq: V_dot}
\end{align}
Using (\ref{eq: r}), (\ref{eq: Nd_bound}), (\ref{eq: N tilde bound}),
the inequality $\dot{\tau}_{s}<1$ and Young's Inequality to show
that $\left\Vert e_{1}^{T}e_{2}\right\Vert \leq\frac{1}{2}\left\Vert e_{1}\right\Vert ^{2}+\frac{1}{2}\left\Vert e_{2}\right\Vert ^{2}$,
$\left\Vert e_{2}^{T}e_{u}\right\Vert \leq\frac{1}{2}\left\Vert e_{2}\right\Vert ^{2}+\frac{1}{2}\left\Vert e_{u}\right\Vert ^{2}$
and $\left\Vert r\right\Vert \rho_{2}\left(\left\Vert z_{\tau_{s}}\right\Vert \right)\left\Vert z_{\tau_{s}}\right\Vert \leq\frac{k_{s}}{2}\left\Vert r\right\Vert ^{2}+\frac{1}{2k_{s}}\rho_{2}^{2}\left(\left\Vert z_{\tau_{s}}\right\Vert \right)\left\Vert z_{\tau_{s}}\right\Vert ^{2}$,
the expression in (\ref{eq: V_dot}) can be upper bounded as 
\begin{align}
\dot{V} & \leq-\alpha_{1}\left\Vert e_{1}\right\Vert ^{2}-\alpha_{2}\left\Vert e_{2}\right\Vert ^{2}-\left(\frac{k_{s}}{2}+1\right)\left\Vert r\right\Vert ^{2}\nonumber \\
 & +\frac{1}{2}\left\Vert e_{1}\right\Vert ^{2}+\left\Vert e_{2}\right\Vert ^{2}+\frac{1}{2}\left\Vert e_{u}\right\Vert ^{2}+\left\Vert r\right\Vert \rho_{1}\left(\left\Vert z\right\Vert \right)\left\Vert z\right\Vert \nonumber \\
 & +\frac{1}{2k_{s}}\rho_{2}^{2}\left(\left\Vert z_{\tau_{s}}\right\Vert \right)\left\Vert z_{\tau_{s}}\right\Vert ^{2}-\frac{\gamma_{1}\left(1-\dot{\tau}_{s}\right)}{2k_{s}}\rho_{2}^{2}\left(\left\Vert z_{\tau_{s}}\right\Vert \right)\left\Vert z_{\tau_{s}}\right\Vert ^{2}\nonumber \\
 & +\varphi_{i_{1}}\left(1+\omega\right)\left(k_{s}+1\right)^{2}\left\Vert r\right\Vert ^{2}+\left\Vert r\right\Vert \zeta_{N_{d1}}\nonumber \\
 & -\omega\left(1-\dot{\tau}_{i}\right)\int_{t-\tau_{i}}^{t}\left\Vert \dot{u}\left(\theta\right)\right\Vert ^{2}d\theta\nonumber \\
 & +\left(\frac{\gamma_{1}}{2k_{s}}+\frac{\gamma_{2}}{k_{s}}\tau_{s}\right)\rho_{2}^{2}\left(\left\Vert z\right\Vert \right)\left\Vert z\right\Vert ^{2}\nonumber \\
 & -\frac{\gamma_{2}}{k_{s}}\left(1-\dot{\tau}_{s}\right)\int_{t-\tau_{s}}^{t}\rho_{2}^{2}\left(\left\Vert z\left(\theta\right)\right\Vert \right)\left\Vert z\left(\theta\right)\right\Vert ^{2}d\theta.\label{eq: V_dot-2}
\end{align}
Completing the squares for $r$, utilizing the inequalities 
\begin{gather*}
\left\Vert e_{u}\right\Vert ^{2}\leq\tau_{i}\int_{t-\tau_{i}}^{t}\left\Vert \dot{u}\left(\theta\right)\right\Vert ^{2}d\theta,\\
\begin{alignedat}{1}-\int_{t-\tau_{i}}^{t}\left\Vert \dot{u}\left(\theta\right)\right\Vert ^{2}d\theta & \leq\\
-\frac{1}{\tau_{i}}\int_{t-\tau_{i}}^{t}\left(\int_{s}^{t}\left\Vert \dot{u}\left(\theta\right)\right\Vert ^{2}d\theta\right)ds & =-\frac{Q_{LK}}{\omega\tau_{i}},
\end{alignedat}
\\
\begin{aligned}-\int_{t-\tau_{s}}^{t}\rho_{2}^{2}\left(\left\Vert z\left(\theta\right)\right\Vert \right)\left\Vert z\left(\theta\right)\right\Vert ^{2}d\theta & \leq\\
-\frac{1}{\tau_{s}}\int_{t-\tau_{s}}^{t}\left(\int_{s}^{t}\rho_{2}^{2}\left(\left\Vert z\left(\theta\right)\right\Vert \right)\left\Vert z\left(\theta\right)\right\Vert ^{2}d\theta\right)ds & =\\
-\frac{k_{s}S_{LK}}{\gamma_{2}\tau_{s}},
\end{aligned}
\end{gather*}
and (\ref{eq:DelayCond2}), (\ref{eq: PLK}) and (\ref{eq: RLK}),
(\ref{eq: V_dot-2}) can be rewritten as
\begin{align}
\dot{V} & \leq-\frac{\alpha_{1}}{2}\left\Vert e_{1}\right\Vert ^{2}-\frac{\alpha_{2}}{2}\left\Vert e_{2}\right\Vert ^{2}-\left\Vert r\right\Vert ^{2}-\frac{\omega\left(1-\dot{\tau}_{i}\right)}{6\tau_{i}}\left\Vert e_{u}\right\Vert ^{2}\nonumber \\
 & -\left(\frac{\alpha_{1}}{2}-\frac{1}{2}\right)\left\Vert e_{1}\right\Vert ^{2}-\left(\frac{\alpha_{2}}{2}-1\right)\left\Vert e_{2}\right\Vert ^{2}\nonumber \\
 & -\left(\frac{k_{s}}{6}-\varphi_{i_{1}}\left(1+\omega\right)\left(k_{s}+1\right)^{2}\right)\left\Vert r\right\Vert ^{2}\nonumber \\
 & -\left(\frac{\omega\left(1-\dot{\tau}_{i}\right)}{6\tau_{i}}-\frac{1}{2}\right)\left\Vert e_{u}\right\Vert ^{2}\nonumber \\
 & -\left(\frac{\gamma_{1}\left(1-\dot{\tau}_{s}\right)}{2k_{s}}-\frac{1}{2k_{s}}\right)\rho_{2}^{2}\left(\left\Vert z_{\tau_{s}}\right\Vert \right)\left\Vert z_{\tau_{s}}\right\Vert ^{2}\nonumber \\
 & +\frac{1}{2k_{s}}\left(3\rho_{1}^{2}\left(\left\Vert z\right\Vert \right)+\left(\gamma_{1}+2\gamma_{2}\varphi_{s_{1}}\right)\rho_{2}^{2}\left(\left\Vert z\right\Vert \right)\right)\left\Vert z\right\Vert ^{2}\nonumber \\
 & -\frac{\omega\left(1-\dot{\tau}_{i}\right)}{3\varphi_{i_{1}}}P_{LK}-\frac{\left(1-\dot{\tau}_{i}\right)}{3\tau_{i}}Q_{LK}-\frac{\gamma_{2}\left(1-\dot{\tau}_{s}\right)}{\gamma_{1}}R_{LK}\nonumber \\
 & -\frac{\left(1-\dot{\tau}_{s}\right)}{2\tau_{s}}S_{LK}+\frac{3\zeta_{N_{d1}}^{2}}{2k_{s}}.\label{eq: V_dot-3}
\end{align}
If the conditions in (\ref{eq: gain cond}) are satisfied, based on
the inequalities $\left\Vert z\right\Vert ^{2}\geq\left\Vert \acute{z}\right\Vert ^{2}$
and $\left\Vert z\right\Vert \leq\left\Vert y\right\Vert $, the expression
in (\ref{eq: V_dot-3}) reduces to

\begin{align}
\dot{V} & \leq-\sigma\left\Vert \acute{z}\right\Vert ^{2}-\frac{\omega\left(1-\varphi_{i_{2}}\right)}{3\varphi_{i_{1}}}P_{LK}-\frac{\left(1-\varphi_{i_{2}}\right)}{3\varphi_{i_{1}}}Q_{LK}\nonumber \\
 & -\frac{\gamma_{2}\left(1-\varphi_{s_{2}}\right)}{\gamma_{1}}R_{LK}-\frac{\left(1-\varphi_{s_{2}}\right)}{2\varphi_{s_{1}}}S_{LK}+\frac{3\zeta_{N_{d1}}^{2}}{2k_{s}},\nonumber \\
 & \leq-\delta\left\Vert y\right\Vert ^{2},\:\forall\left\Vert y\right\Vert \geq\sqrt{\frac{3\zeta_{N_{d1}}^{2}}{2k_{s}\delta}},\label{eq: V_dot-4}
\end{align}
provided $y\in\mathscr{D}$, where $\rho\left(\left\Vert z\right\Vert \right)$,
$\sigma$, and $\delta$ were introduced in (\ref{eq:rho}), (\ref{eq: sigma})
and (\ref{eq:delta}). Using (\ref{eq:DelayCond2}), (\ref{eq: V bound}),
and (\ref{eq: V_dot-4}), Theorem 4.18 in \cite{Khalil2002} can be
invoked to conclude that $y$ is uniformly ultimately bounded in the
sense that $\lim\sup_{t\to\infty}\left\Vert y\right\Vert \leq\sqrt{\frac{3\zeta_{N_{d1}}^{2}}{k_{s}\delta}}$,
provided $y\left(t_{0}\right)\in\mathcal{S}_{\mathscr{D}}$. 

Since $e_{1},e_{2},r\in\mathcal{L}_{\infty}$, from (\ref{eq: open loop}),
$u\in\mathcal{L}_{\infty}$, which implies $u_{\tau i}\in\mathcal{L}_{\infty},$
and hence, $e_{u}\in\mathcal{L}_{\infty}$. The closed-loop error
system can then be used to conclude that the remaining signals are
bounded.\end{IEEEproof}
\begin{rem}
\label{rem:rho}If the system dynamics are such that $\left\Vert \tilde{N}\right\Vert $
is linear in $\left\Vert z\right\Vert $, then the function $\rho$
can be selected to be a constant, i.e., $\rho\left(\left\Vert z\right\Vert \right)=\overline{\rho},\:\forall z\in\mathbb{R}^{4n}$
for some known $\overline{\rho}>0$. In this case, the gain condition
in (\ref{eq:DelayCond2}) reduces to 
\begin{equation}
k_{s}>\frac{\overline{\rho}}{2\sigma},\label{eq:DelayCond3}
\end{equation}
and the result is global in the sense that $\mathscr{D}=\mathcal{S}_{\mathscr{D}}=\mathbb{R}^{3n+4}$.
\end{rem}

\section{Conclusion}

This paper presents a robust controller for uncertain nonlinear systems
which include simultaneous time-varying state and input delays, as
well as sufficiently smooth additive bounded disturbances. The controller
utilizes a robust design approach to compensate for the unknown state
delays coupled with an error system structure that provides a delay-free
open-loop error system. The controller and LK functionals guarantee
uniformly ultimately bounded tracking provided the rates of the delays
are sufficiently slow. The control development can be applied when
there is uncertainty in the system dynamics and when the state delay
is unknown; however, the controller is based on the assumption that
the time-varying input delay is known. Simulation results point to
the possibility that different control or analysis methods could be
developed to eliminate the assumption that the input delay is known.
That is, perhaps the interval of previous control values could somehow
be designed big enough to provide predictive properties despite uncertainty
in the input delay.\bibliographystyle{IEEEtran}
\bibliography{ncr,master,encr}

\end{document}